\documentclass[aps,twocolumn,footinbib,floatfix,
superscriptaddress,
longbibliography,
preprintnumbers]{revtex4-1}

\usepackage{amsmath,amssymb,bm,multirow,float,array,bbold,subfigure}
\usepackage{graphicx,color}
\usepackage[usenames,dvipsnames]{xcolor}
\usepackage{dsfont} 
\usepackage[colorlinks,bookmarksopen,bookmarksnumbered,pdfstartview=FitH,citecolor=SBred, linkcolor=SBred,urlcolor=SBred]{hyperref}
\usepackage{graphicx}
\usepackage{enumitem}

\usepackage{booktabs}
\AtBeginDocument{
\heavyrulewidth=.05em
\lightrulewidth=.05em
\cmidrulewidth=.03em
\belowrulesep=.65ex
\belowbottomsep=0pt
\aboverulesep=.4ex
\abovetopsep=0pt
\cmidrulesep=\doublerulesep
\cmidrulekern=.5em
\defaultaddspace=.5em
}
\definecolor{SBred}{rgb}{0.6471, 0.1098, 0.1882}

\def \refeq#1{(\ref{#1})}

\def \refTab#1{Table~\ref{#1}}

\newcommand{\muEW}{{\mu_\mathrm{ew}}}
\newcommand{\muLow}{{\mu}}

\def\epe{\varepsilon'/\varepsilon}
\def\epsK{\varepsilon_K}

\newcommand{\ord}{\mathcal{O}}

\newcommand{\wT}[1]{\widetilde{#1}}

\begin{document}

\preprint{TUM-HEP-1150/18}
\preprint{AJB-18-6}
\preprint{CP3-18-38}

\title{\boldmath Master formula for \texorpdfstring{$\epe$}{epsilon'/epsilon} beyond the Standard Model}

\author{Jason Aebischer}  
\affiliation{
  Excellence Cluster Universe, Technische Universit\"at M\"unchen,
  Boltzmannstr.~2,
  85748~Garching,
  Germany
}

\author{Christoph Bobeth}  
\affiliation{
  Physik Department, TU M\"unchen,
  James-Franck-Stra{\ss}e,
  85748 Garching,
  Germany
}
\affiliation{
  Excellence Cluster Universe, Technische Universit\"at M\"unchen,
  Boltzmannstr.~2,
  85748~Garching,
  Germany
}

\author{Andrzej J.\ Buras}  
\affiliation{
  TUM Institute for Advanced Study,
  Lichtenbergstr.~2a,
  85748 Garching,
  Germany
}

\author{Jean-Marc G{\'e}rard}  
\affiliation{
  Centre for Cosmology, Particle Physics and Phenomenology (CP3),
  Universit{\'e} catholique de Louvain,
  Chemin du Cyclotron 2,
  1348 Louvain-la-Neuve,
  Belgium
}

\author{David M.\ Straub}  
\affiliation{
  Excellence Cluster Universe, Technische Universit\"at M\"unchen,
  Boltzmannstr.~2,
  85748~Garching,
  Germany
}

\begin{abstract}
\vspace{5mm}
\noindent
We present for the first time a master formula for $\epe$, the ratio probing
direct CP violation in $K \to \pi\pi$ decays, valid in {\em any}
ultraviolet extension of the Standard Model (BSM). The formula makes use
of hadronic matrix elements of
BSM operators calculated recently in the Dual QCD approach and the ones of the
SM operators from lattice QCD. We emphasize the large impact of several scalar
and tensor BSM operators in the context of the emerging $\epe$ anomaly. We have
implemented the results in the open source code flavio.

\vspace{3mm}
\end{abstract}

\maketitle

%
%
%

The non-conservation of the product of parity (P) and charge-conjugation (C)
symmetries in nature, known under the name of CP violation, was established
experimentally for the first time in 1964 via $K\to \pi\pi$
decays~\cite{Christenson:1964fg}. Since then, this fundamental phenomenon has been
confirmed also in other processes in the quark sector and is rather consistently
described by the so-called Cabibbo-Kobayashi-Maskawa (CKM) mixing matrix
\cite{Cabibbo:1963yz, Kobayashi:1973fv} within the Standard Model (SM) of
elementary particle physics. Currently there are experimental efforts to
establish analogous CP violation in the lepton sector.

CP violation proves to be a prerequisite \cite{Sakharov:1967dj} for our present
understanding of matter dominance over anti-matter in the universe. However, the
CP-violating contribution from the CKM matrix in the SM fails to account for
this observation and it remains to be seen whether the CP-violating
contributions in the lepton sector will be able to do so. As direct collider
searches have not yet revealed any presence of new physics, rare processes in
the quark sector remain a good territory to search for new sources of CP
violation. This is especially the case for the kaon physics observables
$\varepsilon$ and $\varepsilon^\prime$, which measure indirect and direct CP
violation in $K^0$-$\bar K^0$ mixing and $K^0$ decay into $\pi\pi$,
respectively.

Recently, there has been a renewed interest in the ratio $\epe$
\cite{Buras:2014sba, Buras:2015yca, Blanke:2015wba, Buras:2015kwd,
  Buras:2016dxz, Buras:2015jaq, Kitahara:2016otd, Endo:2016aws, Endo:2016tnu,
  Cirigliano:2016yhc, Alioli:2017ces, Bobeth:2016llm, Bobeth:2017xry,
  Crivellin:2017gks, Bobeth:2017ecx, Endo:2017ums, Haba:2018byj, Chen:2018ytc,
  Chen:2018vog, Matsuzaki:2018jui, Haba:2018rzf}, due to hints for a significant
tension between measurements and the SM prediction from the RBC-UKQCD lattice
collaboration \cite{Bai:2015nea, Blum:2015ywa} and the Dual QCD approach (DQCD)
\cite{Buras:2015xba, Buras:2016fys}. While on the experimental side the world
average from the NA48 \cite{Batley:2002gn} and KTeV~\cite{AlaviHarati:2002ye,
  Abouzaid:2010ny} collaborations reads
\begin{align}
  \label{EXP}
  (\epe)_\text{exp} &
  = (16.6 \pm 2.3) \times 10^{-4} \,,
\end{align}
the lattice collaboration \cite{Bai:2015nea, Blum:2015ywa} and the NLO analyses
in \cite{Buras:2015yba, Kitahara:2016nld} based on their results find
$(\epe)_{\text{SM}}$ in the ballpark of $(1-2) \times 10^{-4}$, that is by one
order of magnitude below the data, but with an error in the ballpark of
$5\times 10^{-4}$.  An independent analysis based on hadronic matrix elements
from DQCD \cite{Buras:2015xba, Buras:2016fys} gives a strong support to these
values and moreover provides an \textit{upper bound} on $(\epe)_{\text{SM}}$ in
the ballpark of $6\times 10^{-4}$.
A different view has been expressed in \cite{Gisbert:2017vvj} where, using ideas
from chiral perturbation theory but going beyond it, the
authors find $(\epe)_{\text{SM}} = (15 \pm 7) \times 10^{-4}$ in agreement with
the data, albeit with a large uncertainty.

The results from RBC-UKQCD and DQCD motivated several authors to look for
various extensions of the SM which could bring the theory to agree with
data. For a recent review see \cite{Buras:2018wmb}.  In all the models studied
to date, the rescue comes from the modification of the Wilson coefficient of the
dominant electroweak left-right (LR) penguin operator $Q_8$, but also solutions
through a modified contribution of the dominant QCD LR penguin operator $Q_6$
could be considered \cite{Buras:2015jaq}. However, in generic BSM scenarios,
also operators not present in the SM could play an important role. The very
recent calculation of the $K\to\pi\pi$ hadronic matrix elements of all BSM
four-quark operators, in particular scalar and tensor operators in DQCD
\cite{Aebischer:2018rrz} and the one of the chromo-magnetic operator by the ETM
lattice collaboration \cite{Constantinou:2017sgv} and in DQCD
\cite{Buras:2018evv}, allow for the first time the study of $\epe$ in an
arbitrary extension of the SM.  While the matrix element of the chromo-magnetic
operator has been found to be much smaller than previously expected, the values
of the BSM matrix elements of scalar and tensor operators are found to be in the
ballpark of the ones of $Q_8$, the dominant electroweak penguin operator in the
SM. Consequently, these operators could help in the explanation of the emerging
$\epe$ anomaly.

As far as short-distance contributions encoded in the Wilson coefficients are
concerned, they have been known for the SM operators already for 25 years at the
NLO level \cite{Buras:1991jm, Buras:1992tc, Buras:1992zv, Ciuchini:1992tj,
  Buras:1993dy, Ciuchini:1993vr} and for the BSM operators two-loop anomalous
dimensions have been known \cite{Ciuchini:1997bw, Buras:2000if} for almost two
decades. First steps towards the NNLO predictions for $\epe$ have been made in
\cite{Bobeth:1999mk, Buras:1999st, Gorbahn:2004my,Brod:2010mj} and the complete
NNLO result should be available soon \cite{Cerda-Sevilla:2016yzo}.

\bigskip

Having all these ingredients from long-distance and short-distance contributions
at hand, we are in the position to present for the first time a master formula
for $\epe$ that can be applied to any ultraviolet extension of the SM.
Neglecting isospin breaking corrections, $\epe$ can be written as
\begin{align}
  \label{eq:epe-formula}
  \left(\frac{\varepsilon'}{\varepsilon}\right)_\text{th} &
  = -\frac{\omega}{\sqrt{2}|\epsK|}
    \left[ \frac{\text{Im}A_0}{\text{Re}A_0}
         - \frac{\text{Im}A_2}{\text{Re}A_2} \right]\,,
\end{align}
where $\omega = {\text{Re}A_2}/{\text{Re}A_0}$ and $A_{0,2}$ are the
$K\to\pi\pi$ isospin amplitudes,
\begin{align}
  A_{0,2} &
  = \big\langle (\pi\pi)_{I=0,2}\, \big|\; \mathcal{H}_{\Delta S = 1}^{(3)}
  \;\big|\, K^0 \big\rangle \,.
\end{align}
Isospin breaking corrections have been considered in~\cite{Cirigliano:2003gt,
  Cirigliano:2003nn}. These corrections will affect only
the $A_0$ contributions that are suppressed by $\omega \sim 1/22$. They can only
be relevant in NP  scenarios in which, similar to the case of the SM, the Wilson
coefficients of the operators contributing to $A_0$ are by more than one order
of magnitude larger than those relevant for the $A_2$ amplitude.
Here $\mathcal{H}_{\Delta S = 1}^{(3)}$ denotes the
$\Delta S=1$ effective Hamiltonian with only the three lightest quarks
($q=u,d,s$) being dynamical, obtained by decoupling the heavy $W^\pm$, $Z^0$,
and $h^0$ bosons and the top quark at the electroweak scale $\muEW \sim m_W$ and
the bottom and charm quarks at their respective mass thresholds
\cite{Buchalla:1995vs}.

Assuming that no particles beyond the SM ones with mass below the electroweak
scale exist, any BSM effect is encoded in the Wilson coefficients of the most
general $\Delta S = 1$ dimension-six effective Hamiltonian.  The values of the
Wilson coefficients $C_i(\muEW)$ in this effective Hamiltonian at the electroweak
scale with $N_f = 5$ active quark flavours,
\begin{align}
  \label{eq:DS1-Hamiltonian}
  \mathcal{H}_{\Delta S = 1}^{(5)} &
  = - \mathcal{N}_{\Delta S = 1} \sum_i C_i \, O_i \,,
\end{align}
are connected to those of $\mathcal{H}_{\Delta S = 1}^{(3)}$, entering $\epe$,
by the usual QCD and QED renormalization group (RG) evolution.  In full
generality, three classes of operators can contribute, directly or via RG
mixing, to $K\to\pi\pi$ decays:

\paragraph{four-quark operators:}

\begin{align}
  \label{eq:DS1-psi4-col1}
  O_{XAB}^q &
  = (\bar s^i \Gamma_X P_A d^i) (\bar q^j \Gamma_X P_B q^j) \,,
\\
  \label{eq:DS1-psi4-col8}
  \widetilde{O}_{XAB}^q &
  = (\bar s^i \Gamma_X P_A d^j) (\bar q^j \Gamma_X P_B q^i) \,,
\end{align}

\paragraph{electro- and chromo-magnetic dipole operators:}

\begin{align}
  \label{eq:DS1-dipole-QED}
  O_{7\gamma}^{(\prime)} &
  = m_s(\bar s \sigma^{\mu\nu} P_{L(R)} d) F_{\mu\nu} \,,
\\
  \label{eq:DS1-dipole-QCD}
  O_{8g}^{(\prime)}      &
  = m_s(\bar s \sigma^{\mu\nu} T^a P_{L(R)} d) G^a_{\mu\nu} \,,
\end{align}

\paragraph{semi-leptonic operators:}

\begin{align}
  O_{XAB}^\ell &
  = (\bar s\, \Gamma_X P_A d) (\bar \ell\, \Gamma_X P_B \ell) \,.
\end{align}
Here $i,j$ are colour indices, $A,B=L,R$, and $X=S,V,T$ with $\Gamma_S=1$,
$\Gamma_V=\gamma^\mu$, $\Gamma_T=\sigma^{\mu\nu}$ \footnote{For $\Gamma_T$ there
  is only $P_A = P_B$ in four dimensions but not $P_A \neq P_B$.}. Throughout it
is sufficient to consider the case $A = L$, whereas results for the case $A = R$
follow analogously due to parity conservation of QCD and QED. We will choose the
overall normalization factor $\mathcal{N}_{\Delta S = 1}$ below such that the
coefficients $C_i$ are dimensionless.

In the following, we will neglect the electro-magnetic dipole and semi-leptonic
operators, which only enter through small QED effects.  This leaves 40
four-quark operators for $N_f = 5$ and one chromo-magnetic dipole operator of a
given chirality which have to be considered at the electroweak scale.  A
detailed renormalization group analysis of these operators, model independently
and in the context of the Standard Model effective field theory (SMEFT), is
performed in \cite{Aebischer:2018csl}.  The goal of the present letter is to
provide the central result of \cite{Aebischer:2018csl} and~\cite{Aebischer:2018rrz},
the master formula for $\epe$, in a form that could be used by any model builder
or phenomenologist right away without getting involved with the technical
intricacies of these analyses.

Writing
\begin{align}
  \label{eq:mastertotal}
  \left(\frac{\varepsilon'}{\varepsilon}\right) &
  = \left(\frac{\varepsilon'}{\varepsilon}\right)_\text{SM}
  + \left(\frac{\varepsilon'}{\varepsilon}\right)_\text{BSM},
\end{align}
our formula allows to calculate automatically $(\epe)_\text{BSM}$ once the
Wilson coefficients of all contributing operators are known at the electroweak
scale $\muEW$. It reads as follows:
\begin{align}
  \label{eq:master}
  \left(\frac{\varepsilon'}{\varepsilon}\right)_\text{BSM} &
  = \sum_i  P_i(\muEW) ~\text{Im}\left[ C_i(\muEW)-C^\prime_i(\muEW)\right],
\end{align}
where
\begin{equation}\label{eq:master2}
  P_i(\muEW)  = \sum_{j} \sum_{I=0,2} p_{ij}^{(I)}(\muEW, \muLow)
  \,\left[\frac{\langle Q_j (\muLow)\rangle_I}{\text{GeV}^3}\right]\,.
\end{equation}

\begin{table}
 \footnotesize
\renewcommand{\arraystretch}{1.2}
\begin{tabular}{clrrc}
\hline
  class & $O_i$       & $P_i$ & $\frac{\Lambda}{\text{TeV}}$ & \textsc{smeft}
\\
\hline
  \multirow{18}{*}{A)}
& $O_{VLL}^u = (\bar s^i \gamma_\mu P_L d^i)(\bar u^j \gamma^\mu P_L u^j)$                  &  $-4.3 \pm 1.0$    &   65 &  \checkmark \\
& $O_{VLR}^u = (\bar s^i \gamma_\mu P_L d^i)(\bar u^j \gamma^\mu P_R u^j)$                  &  $-126 \pm 10$      &  354 &  \checkmark \\
& $\wT{O}_{VLL}^u = (\bar s^i \gamma_\mu P_L d^j)(\bar u^j \gamma^\mu P_L u^i)$             &  $1.5 \pm 1.7$     &   38 &  \checkmark \\
& $\wT{O}_{VLR}^u = (\bar s^i \gamma_\mu P_L d^j)(\bar u^j \gamma^\mu P_R u^i)$             &  $-436 \pm 35$     &  659 &  \checkmark \\[0.1cm]
& $O_{VLL}^d = (\bar s^i \gamma_\mu P_L d^i)(\bar d^j \gamma^\mu P_L d^j)$                  &  $2.3 \pm 0.5$     &   48 &  \checkmark \\
& $O_{VLR}^d = (\bar s^i \gamma_\mu P_L d^i)(\bar d^j \gamma^\mu P_R d^j)$                  &  $123 \pm 10$      &  350 &  \checkmark \\
& $O_{SLR}^d = (\bar s^i P_L d^i)(\bar d^j P_R d^j)$                                        &  $214 \pm 19$ &  462 &  \checkmark \\[0.1cm]
& $O_{VLL}^s = (\bar s^i \gamma_\mu P_L d^i)(\bar s^j \gamma^\mu P_L s^j)$                  &  $-0.4 \pm 0.1$    &   18 &  \checkmark \\
& $O_{VLR}^s = (\bar s^i \gamma_\mu P_L d^i)(\bar s^j \gamma^\mu P_R s^j)$                  &  $-0.32 \pm 0.05$  &   17 &  \checkmark \\
& $O_{SLR}^s = (\bar s^i P_L d^i)(\bar s^j P_R s^j)$                                        &  $0.0 \pm 0.1$    &   6  &  \checkmark \\[0.1cm]
& $O_{VLL}^c = (\bar s^i \gamma_\mu P_L d^i)(\bar c^j \gamma^\mu P_L c^j)$                  &  $0.7 \pm 0.1$     &   25 &  \checkmark \\
& $O_{VLR}^c = (\bar s^i \gamma_\mu P_L d^i)(\bar c^j \gamma^\mu P_R c^j)$                  &  $0.7 \pm 0.1$     &   26 &  \checkmark \\
& $\wT{O}_{VLL}^c = (\bar s^i \gamma_\mu P_L d^j)(\bar c^j \gamma^\mu P_L c^i)$             &  $0.2 \pm 0.2$     &   13 &  \checkmark \\
& $\wT{O}_{VLR}^c = (\bar s^i \gamma_\mu P_L d^j)(\bar c^j \gamma^\mu P_R c^i)$             &  $0.4 \pm 0.2$     &   20 &  \checkmark \\[0.1cm]
& $O_{VLL}^b = (\bar s^i \gamma_\mu P_L d^i)(\bar b^j \gamma^\mu P_L b^j)$                  &  $-0.30 \pm 0.03$  &   17 &  \checkmark \\
& $O_{VLR}^b = (\bar s^i \gamma_\mu P_L d^i)(\bar b^j \gamma^\mu P_R b^j)$                  &  $-0.28 \pm 0.03$  &   16 &  \checkmark \\
& $\wT{O}_{VLL}^b = (\bar s^i \gamma_\mu P_L d^j)(\bar b^j \gamma^\mu P_L b^i)$             &  $0.0 \pm 0.1$    &    4 &  \checkmark \\
& $\wT{O}_{VLR}^b = (\bar s^i \gamma_\mu P_L d^j)(\bar b^j \gamma^\mu P_R b^i)$             &  $-0.1 \pm 0.1$    &    8 &  \checkmark \\
\hline
  \multirow{11}{*}{B)}
& $O_{8g} \;\;\, = m_s (\bar s \sigma^{\mu\nu} T^a P_L d)\, G^{a}_{\mu\nu}$                 &  $-0.35 \pm 0.12$    &   18 &  \checkmark \\[0.1cm]
& $O_{SLL}^s = (\bar s^i P_L d^i)(\bar s^j P_L s^j)$                                        &  $0.05 \pm 0.02$  &    7 &  \\
& $O_{TLL}^s = (\bar s^i \sigma_{\mu\nu} P_L d^i)(\bar s^j \sigma^{\mu\nu} P_L s^j)$        &  $-0.14 \pm 0.05$   &   12 &  \\[0.1cm]
& $O_{SLL}^c = (\bar s^i P_L d^i)(\bar c^j P_L c^j)$                                        &  $-0.26 \pm 0.09$     &   16 &  \checkmark \\
& $O_{TLL}^c = (\bar s^i \sigma_{\mu\nu} P_L d^i)(\bar c^j \sigma^{\mu\nu} P_L c^j)$        &  $-0.15 \pm 0.05$   &   12 &  \checkmark \\
& $\wT{O}_{SLL}^c = (\bar s^i P_L d^j)(\bar c^j P_L c^i)$                                   &  $-0.23 \pm 0.07$     &   15 &  \checkmark \\
& $\wT{O}_{TLL}^c = (\bar s^i \sigma_{\mu\nu} P_L d^j)(\bar c^j \sigma^{\mu\nu} P_L c^i)$   &  $-5.9 \pm 1.9$     &   76 &  \checkmark \\[0.1cm]
& $O_{SLL}^b = (\bar s^i P_L d^i)(\bar b^j P_L b^j)$                                        &  $-0.35 \pm 0.12$     &   18 &    \\
& $O_{TLL}^b = (\bar s^i \sigma_{\mu\nu} P_L d^i)(\bar b^j \sigma^{\mu\nu} P_L b^j)$        &  $-0.11 \pm 0.03$   &   10 &    \\
& $\wT{O}_{SLL}^b = (\bar s^i P_L d^j)(\bar b^j P_L b^i)$                                   &  $-0.34 \pm 0.11$     &   18 &    \\
& $\wT{O}_{TLL}^b = (\bar s^i \sigma_{\mu\nu} P_L d^j)(\bar b^j \sigma^{\mu\nu} P_L b^i)$   &  $-13.4 \pm 4.5$    &  115 &    \\
\hline
  \multirow{4}{*}{C)}
& $O_{SLL}^u = (\bar s^i P_L d^i)(\bar u^j P_L u^j)$                                        &  $74 \pm 16$      &  272 &  \checkmark \\
& $O_{TLL}^u = (\bar s^i \sigma_{\mu\nu} P_L d^i)(\bar u^j \sigma^{\mu\nu} P_L u^j)$        &  $-162 \pm 36$      &  402 &  \checkmark \\
& $\wT{O}_{SLL}^u = (\bar s^i P_L d^j)(\bar u^j P_L u^i)$                                   &  $-15.6 \pm 3.3$    &  124 &  \checkmark \\
& $\wT{O}_{TLL}^u = (\bar s^i \sigma_{\mu\nu} P_L d^j)(\bar u^j \sigma^{\mu\nu} P_L u^i)$   &  $-509 \pm 108$     &  713 &  \checkmark \\
\hline
  \multirow{2}{*}{D)}
& $O_{SLL}^d = (\bar s^i P_L d^i)(\bar d^j P_L d^j)$                                        &  $-87 \pm 16$       &  295 &  \\
& $O_{TLL}^d = (\bar s^i \sigma_{\mu\nu} P_L d^i)(\bar d^j \sigma^{\mu\nu} P_L d^j)$        &  $191 \pm 35$     &  436 &  \\
\hline
  \multirow{2}{*}{E)}
& $O_{SLR}^u = (\bar s^i P_L d^i)(\bar u^j P_R u^j)$                                        &  $-266 \pm 21$      &  515 &  \\
& $\wT{O}_{SLR}^u = (\bar s^i P_L d^j)(\bar u^j P_R u^i)$                                   &  $-60 \pm 5$        &  244 &  \checkmark \\
\hline
\end{tabular}
\renewcommand{\arraystretch}{1.2}
\caption{\label{tab:P_i}
  Table of $\Delta S = 1$ operators contributing to $(\epe)_\text{BSM}$ with
  coefficients $P_i(\muEW)$ for $\muEW=160$\,GeV, and corresponding suppression
  scales. The Hamiltonian is normalized as
  $\mathcal{H}_{\Delta S = 1}^{(5)} = - \sum_i \frac{C_i(\muEW)}{(1\,\text{TeV})^2} \, O_i \,$.
}
\end{table}

The sum in \eqref{eq:master2} extends over the Wilson coefficients $C_i$ of the
linearly independent four-quark and chromo-magnetic dipole operators listed in
\refTab{tab:P_i}.  The $C_i'$ are the Wilson coefficients of the corresponding
chirality-flipped operators obtained by replacing $P_L\leftrightarrow P_R$.  The
relative minus sign accounts for the fact that their $K\to\pi\pi$ matrix
elements differ by a sign.  Among the contributing operators are also operators
present already in the SM but their Wilson coefficients in \refeq{eq:master}
include only BSM contributions.

The dimensionless coefficients $p_{ij}^{(I)}(\muEW,\muLow)$ include the QCD and
QED RG evolution from $\muEW$ to $\muLow \sim \ord(1\,\text{GeV})$ for each
Wilson coefficient as well as the relative suppression of the contributions to
the $I=0$ amplitude due to ${\text{Re}A_2} / {\text{Re}A_0}\ll 1$ for the matrix
elements $\langle Q_j (\muLow) \rangle_I$ of all the operators $Q_j$ present at
the low-energy scale. The index $j$ includes also $i$ so that the effect of
self-mixing is included. We refer the reader to \cite{Aebischer:2018csl}
for the numerical values of the $p_{ij}^{(I)}(\muEW,\muLow)$ and $\langle Q_j
(\muLow) \rangle_I$ for our choice of the set of $Q_j$. The details given their
allow to easily account for future updates of the matrix elements.
The $P_i(\muEW)$ do not depend on $\muLow$ to the
considered order, because the $\muLow$-dependence cancels between matrix
elements and the RG evolution operator.  Moreover, it should be emphasized that
their values are {\em model-independent} and depend only on the SM dynamics
below the electroweak scale, which includes short distance contributions down to
$\muLow$ and the long distance contributions represented by the hadronic matrix
elements. The BSM dependence enters our master formula in (\ref{eq:master}) {\em
  only} through the Wilson coefficients $C_i(\muEW)$ and
$C^\prime_i(\muEW)$. That is, even if a given $P_i$ is non-zero, the fate of its
contribution depends on the difference of these two coefficients. In particular,
in models with exact left-right symmetry this contribution vanishes as first
pointed out in \cite{Branco:1982wp}.

The numerical values of the $P_i(\muEW)$ are collected in \refTab{tab:P_i} for
\begin{align}
 \muEW&=160\,\text{GeV} \,,&
 \mathcal{N}_{\Delta S = 1}&=(1\,\text{TeV})^{-2} \,.
\end{align}
They have been calculated with the flavio package~\cite{Straub:2018kue}, where we have
implemented general BSM contributions to $\epe$.  As seen in (\ref{eq:master2}),
the $P_i$ depend on the hadronic matrix elements
$\langle Q_j (\muLow) \rangle_I$ and the RG evolution factors
$p_{ij}^{(I)}(\muEW, \muLow)$.  The numerical values of the hadronic matrix
elements rely on lattice QCD in the case of SM operators \cite{Bai:2015nea,
  Blum:2015ywa} and on results for scalar and tensor operators obtained in DQCD
\cite{Aebischer:2018rrz}.  The matrix element of the chromo-magnetic dipole
operator is from \cite{Constantinou:2017sgv} and \cite{Buras:2018evv} which
agree with each other.

The operators in \refTab{tab:P_i} have been grouped into five distinct classes.

{\bf Class A:} All hadronic matrix elements can be expressed in terms of
the ones of SM operators calculated by lattice QCD \cite{Bai:2015nea,
  Blum:2015ywa}.

{\bf Class B:} All operators contribute only through RG mixing into the
chromo-magnetic operator $O_{8g}$ so that only one hadronic matrix element is
involved and taken from \cite{Constantinou:2017sgv,Buras:2018evv}.

{\bf Class C:} RLRL type operators with flavour $(\bar sd)(\bar uu)$ that
contribute via BSM matrix elements \cite{Aebischer:2018rrz} or by generating
the chromo-magnetic dipole matrix element \cite{Constantinou:2017sgv,
Buras:2018evv} through mixing.

{\bf Class D:} RLRL type operators with flavour $(\bar sd)(\bar dd)$ that
contribute via BSM matrix elements \cite{Aebischer:2018rrz} or the
chromo-magnetic dipole matrix element \cite{Constantinou:2017sgv,
  Buras:2018evv}.

{\bf Class E:} RLLR type operators with flavour $(\bar sd)(\bar uu)$ that
contribute exclusively via BSM matrix elements \cite{Aebischer:2018rrz}.

Besides the $P_i$, we provide in the next-to-last column of \refTab{tab:P_i} the
suppression scale $\Lambda$ that would generate $(\epe)_\text{BSM}=10^{-3}$ for
$C_i=(1\,\text{TeV})^2/\Lambda^2$. It gives an indication of the maximal scale
probed by $\epe$ for any given operator.

Among the 40 four-quark operators present in $\mathcal{H}_{\Delta S = 1}^{(5)}$,
four have been omitted in \refTab{tab:P_i}, namely $O_{SLR}^{b,c}$ and
$\widetilde{O}_{SLR}^{b,c}$, since they neither contribute directly nor via RG
mixing at the level considered, i.e.\ they have $P_i=0$.

In models with a mass gap above the electroweak scale, $v\ll\Lambda$, where $v$
is the Higgs vacuum expectation value and $\Lambda$ the BSM scale, some of the
operators in \refTab{tab:P_i} are not generated at leading order in an expansion
in $v/\Lambda$. As discussed in more detail in \cite{Aebischer:2018csl}, these
operators violate hypercharge, that is conserved in the SMEFT above the
electroweak scale \footnote{As an exception, the hypercharge
constraint can be avoided for the operator $\tilde{O}_{SLR}^u$, if
in the intermediate SMEFT the dimension-six operator with right-handed
modified $W^\pm$ couplings ($\mathcal{O}_{Hud}$ in the basis of
\cite{Grzadkowski:2010es}) is generated, as for example in a left-right
symmetric UV completion of the SM due to tree-level  $W_L$--$W_R$ mixing.
The $\tilde{O}_{SLR}^u$ is then generated at the electroweak scale
by the tree-level $W^\pm$ exchange of a single insertion of $\mathcal{O}_{Hud}$
with a dimension-four SM coupling of $W^\pm$ and quarks.}. In the rightmost
column of \refTab{tab:P_i}, we have
indicated whether the operator can arise from a tree-level matching of SMEFT at
dimension six onto the $\Delta S=1$ effective Hamiltonian
(cf.~\cite{Aebischer:2015fzz,Jenkins:2017jig}).

Inspecting the results in \refTab{tab:P_i}, the following comments are in order.
\begin{itemize}
\item The largest $P_i$ values in Class~A can be traced back to the large values
  of the matrix elements $\langle Q_{7,8}\rangle_2$, the dominant electroweak
  penguin operators in the SM, and the enhancement by $1/\omega\approx 22$ of the
  $I=2$ contributions.
\item The small $P_i$ values in Class~B are the consequence of the fact that
  each one is proportional to $\langle O_{8g} \rangle_0$, which has recently been
  found to be much smaller than previously expected \cite{Constantinou:2017sgv,
    Buras:2018evv}. Moreover, as $\langle O_{8g}\rangle_2=0$, all contributions
  in this class are suppressed by the factor $1/\omega$ relative to
  contributions from other classes.
\item The large $P_i$ values in Classes C and D can be traced back to the large
  hadronic matrix elements of scalar and tensor operators calculated recently in
  \cite{Aebischer:2018rrz}.  Due to the smallness of $\langle O_{8g} \rangle_0$,
  the contribution of the chromo-magnetic dipole operator is negligible.
\item While the operators in Classes D have sizable $P_i$, they violate
  hypercharge as discussed above, so they do not arise in a tree-level matching
  from SMEFT at dimension six.
\item While the $I=0$ matrix elements of the operators in Class E cannot
be expressed in terms of SM ones, the $I=2$ matrix elements can, and the
large $P_i$ values can be traced back to the large SM matrix elements $\langle Q_{7,8}\rangle_2$.
\end{itemize}

Almost all existing BSM analyses of $\epe$ in the literature are based on the
contributions of operators from Class~A or the chromo-magnetic dipole operator.
\refTab{tab:P_i} shows that also other operators, in particular the ones in
Class~C, could be promising to explain the emerging $\epe$ anomaly and can play
an important role in constraining BSM scenarios.

However, in a concrete BSM scenario, the Wilson coefficients with the highest
values of $P_i$ could vanish or be suppressed by small couplings.  Moreover,
additional constraints on Wilson coefficients can come in SMEFT and from
correlations with other observables as discussed in more detail in
\cite{Aebischer:2018csl}.

\bigskip

Next, we would like to comment on the accuracy of the values of the $P_i$ listed
in \refTab{tab:P_i}.  As far as short distance contributions to the $P_i$ are
concerned, they have been calculated in the leading logarithmic approximation to
RG improved perturbation theory using the results of
\cite{Aebischer:2017gaw,Jenkins:2017dyc,Aebischer:2017ugx,flavio,Aebischer:2018bkb}.
Although the inclusion of next-to-leading corrections is possible already now, such
contributions are renormalization scheme dependent and can only be cancelled by
the one of the hadronic matrix elements. While in the case of SM operators this
dependence has been included in the DQCD calculations in \cite{Buras:2014maa},
much more work still has to be done in the case of BSM operators.

The uncertainties from the matrix elements depend on the operator classes in
\refTab{tab:P_i}. In the $P_i$ column, we have given the uncertainties obtained from
varying the individual matrix elements, assuming them to be uncorrelated.
In Class~A, they stem from the lattice matrix elements. Here
we point out that due to the enhancement of the $I=2$ contributions by the
factor $1/\omega\approx22$, the largest $P_i$ are dominated by the $I=2$ matrix
elements, which are known to 5--7\% accuracy from lattice QCD
\cite{Blum:2015ywa}.
For the small $P_i$ in Class~A, in some cases there are cancellations between
contributions from different matrix elements, leading to larger relative
uncertainties.
The matrix elements of four-quark BSM operators entering
Classes~C--E have only been calculated recently in DQCD approach
\cite{Aebischer:2018rrz} and it will still take some time before lattice QCD
will be able to provide results for them. Previous results of DQCD
imply that it is a successful approximation of low-energy QCD and that the
uncertainties in the largest $P_i$ are at most at the level of $20\%$. While
not as precise as ultimate lattice QCD calculations, DQCD offered over many
years an insight in the lattice results and often, like was the case of the
$\Delta I = 1/2$ rule \cite{Bardeen:1986vz} and the parameter $\hat B_K$
\cite{Bardeen:1987vg}, provided results almost three decades before this was
possible with lattice QCD.  The agreement between results from DQCD and
lattice QCD is remarkable, in particular considering the simplicity of the
former approach compared to the sophisticated and computationally demanding
numerical lattice QCD one. The most recent example of this agreement was an
explanation by DQCD of the pattern of values of $B_6^{(1/2)}$ and
$B_8^{(3/2)}$ entering $\epe$ obtained by lattice QCD \cite{Buras:2015xba,
Buras:2016fys} and of the pattern of lattice values for BSM parameters $B_i$
in $K^0$-$\bar K^0$ mixing \cite{Buras:2018lgu}. This should be sufficient for
the exploration of new phenomena responsible for the hinted $\epe$ anomaly.
Similar comments apply to the hadronic matrix element of the chromo-magnetic
dipole operator, entering mainly the $P_i$ in Class~B, that was recently
calculated in DQCD in \cite{Buras:2018evv} and found to be in agreement with the
lattice QCD result from \cite{Constantinou:2017sgv}.
Since the $P_i$ in Class~B only receive a single contribution,
their relative uncertainties mirror the relative uncertainty of the chromo-magnetic
matrix element, that was estimated at 30\% in \cite{Buras:2018evv}.

\bigskip

The usefulness of our master formula is twofold.  First, it opens the road to an
efficient search for BSM scenarios behind the $\epe$ anomaly and through the
values of $P_i$ in \refTab{tab:P_i} indicates which routes could be more
successful than others.  This will play an important role if the $\epe$ anomaly
will be confirmed by more precise lattice QCD calculations.  Second, it allows
to put strong constraints on models with new sources of CP violation, in many
cases probing scales up to hundreds of TeV, as shown in \refTab{tab:P_i}.  Thus,
even if future lattice QCD calculations within the SM will confirm the data on
$\epe$, our master formula will be instrumental in putting strong constraints on
the parameters of a multitude of BSM scenarios.

\bigskip

\paragraph*{Acknowledgments.}

The work of J.~A., C.~B., A.~J.~B., and D.~M.~S.\ is supported by the DFG cluster of
excellence ``Origin and Structure of the Universe''.
We thank Andreas Crivellin for pointing out misprints in \refTab{tab:P_i}.

%
%

\bibliographystyle{apsrev4-1}
\bibliography{refs}

\end{document}